\documentclass[twocolumn,showpacs,preprintnumbers,amsmath,amssymb,superscriptaddress,prl,longbibliography]{revtex4-2}

\usepackage{graphicx}
\usepackage[hidelinks]{hyperref}
\allowdisplaybreaks

\begin{document}

\title{Shear viscosity in interacting two-dimensional Fermi liquids}

\author{Ulf Gran}
\email{ulf.gran@chalmers.se}
\affiliation{Department of Physics, Chalmers University of Technology, 41296 Gothenburg, Sweden}

\author{Eric Nilsson}
\email{nieric@chalmers.se}
\affiliation{Department of Physics, Chalmers University of Technology, 41296 Gothenburg, Sweden}

\author{Johannes Hofmann}
\email{johannes.hofmann@physics.gu.se}
\affiliation{Department of Physics, Gothenburg University, 41296 Gothenburg, Sweden}
\affiliation{Nordita, Stockholm University and KTH Royal Institute of Technology, 10691 Stockholm, Sweden}

\date{\today}

\begin{abstract}
In interaction-dominated two-dimensional electron gases at intermediate temperatures, electron transport is not diffusive as in the conventional Drude picture but instead hydrodynamic. The relevant transport coefficient in this regime is the shear viscosity. 
Here, we develop a numerically exact basis expansion to solve the Fermi liquid equation, and apply it to compute the shear viscosity of the electron gas with screened Coulomb interactions.
Our calculations are valid at all temperatures and in particular describe the response beyond the asymptotic low-temperature  limit, where perturbative approaches exist.
We show that even in this low-temperature limit, 
there is a nonanalytic exchange contribution to the shear viscosity, highlighting the need for a full nonperturbative solution of the Fermi liquid equation. 
We hope that the techniques developed in this work will serve as a platform to determine the response of interacting Fermi liquids.
\end{abstract}

\maketitle

Electron transport in conventional metals is dominated by momentum-relaxing scattering, which is phonon scattering at high temperature or impurity scattering at low temperature. Here, electrons diffuse in the material and the relevant transport coefficient is the conductivity $\sigma$, which links the drift current ${\bf j}$ to an external field \mbox{$- e \nabla\phi$} by Ohm's law, \mbox{${\bf j} = - e \sigma \nabla \phi$}~\cite{ashcroft76,girvin19}. By contrast, recent experiments in ultraclean 2DEG in graphene and GaAs access a regime of collective electron flow that is described by hydrodynamics~\cite{dejong95,buhmann02,bandurin16,crossno16}, \mbox{$\eta \nabla^2 {\bf j} = - e \nabla \phi$}, where the central transport coefficient is the viscosity $\eta$ that is set by electron interactions (which do not relax momentum). Such viscous electron flow gives rise to non-Ohmic velocity profiles that involve electron backflow and vortices~\cite{bandurin16,gooth18,gusev18,bandurin18,berdyugin19,jenkins22}. In addition, recent advances in crystal growth of ultraclean samples~\cite{gupta21} and device fabrication~\cite{keser21} allow high-precision measurements of the shear viscosity, calling for  accurate many-body theories of the interacting electron gas.

A theoretical calculation of the viscosity even on a kinetic level is complicated since the viscous electron gas is (by definition) interaction-dominated and the hydrodynamic transport regime is accessed at temperatures close to $T_F$. Here, a significant broadening of the Fermi surface has occurred and there is inefficient phase-space suppression of electron interactions. The challenge in this regime is the treatment of collisions, which are described by an integral equation, the collision integral, for which usually only approximate approaches (such as the relaxation time approximation) are feasible. For extended Fermi surfaces, one simplification arises when analyzing the low-temperature form of the viscosity to leading-logarithmic order in the temperature~\cite{novikov06,alekseev20}. This dates back to classical works by~\textcite{abrikosov57, abrikosov59} and exploits the fact that excitations are restricted to a small thermal region around the Fermi surface, reducing the relaxation to an effective angular dynamics~\cite{smith89}. An additional approach at low temperatures is to evaluate the shear viscosity spectral function at high frequencies, for which perturbative many-body approaches apply that can be extrapolated to the static limit using viscoelastic correlators~\cite{conti99,principi16,link18}. Moreover, in a separate context of low-temperature quantum critical systems like graphene at charge neutrality, only certain modes need to be included that are enhanced by collinear scattering, allowing for explicit calculations~\cite{mueller09,kiselev19}. However, despite extensive literature on the shear viscosity, a direct calculation of this quantity from Fermi liquid theory beyond low temperatures appears to be regarded as prohibitively complex.

In this work, we provide a numerically exact and rapidly converging calculation of the shear viscosity and its spectral function from Fermi liquid theory valid at all temperatures. To achieve this, we employ a basis expansion method whose efficacy has recently been demonstrated to determine the longest-lived modes of a Fermi liquid~\cite{hofmann22b}. We consider electrons with a parabolic single-particle dispersion \mbox{$\varepsilon_{\bf p} = \hbar^2p^2/2m^*$}, where $m^*$ is the effective mass, that interact via a screened Coulomb interaction \mbox{$V({\bf q}) = 2\pi e^2/(q+q_{\rm TF})$}~\cite{giuliani05}. Here, \mbox{$q_{\rm TF} = 2/a_B$} with Bohr radius \mbox{$a_B = \hbar^2/m^* e^2$}, and the strength of the interaction is set by the dimensionless parameter \mbox{$r_s =1/a_B\sqrt{\pi n}$}, where $n$ is the electron density. This description applies, for example, to doped graphene below the Fermi temperature with negligible thermal interband excitations, and to GaAs, for which \mbox{$r_s = 1.87/\sqrt{n/10^{11}{\rm cm}^{-2}}$}~\cite{giuliani05}.

Formally, the shear viscosity gives a dissipative contribution to the stress tensor that describes the energy dissipation in response to a perpendicular velocity gradient in the electron gas, $\Pi_{xy} = - \eta \partial_y V_x$. This stress tensor is linked to the electron velocities via a constitutive relation 
\begin{equation}
\Pi_{xy} = 2 m^{*} \int \frac{d{\bf p}}{(2\pi)^2} v_x v_y f ,
\end{equation} 
which is an average over the single-particle distribution function $f(t, {\bf r}, {\bf p})$ that depends on a time $t$, position ${\bf r}$, and wave vector ${\bf p}$ (the factor $2$ accounts for the spin degeneracy). This distribution is obtained from a solution of the Fermi liquid kinetic equation
\begin{equation}
\begin{split}\label{eq:kinetic}
\biggl[\frac{\partial}{\partial t} + \dot{\bf r} \frac{\partial}{\partial {\bf r}} + \dot{\bf p} \frac{\partial}{\partial {\bf p}}\biggr]f &= {\cal J}\{f\} ,
\end{split}
\end{equation}
where $\dot{\bf r} = \hbar {\bf p}/m^*$ is the quasiparticle velocity and in the absence of external forces, \mbox{$\dot{\bf p}=0$}. Here, ${\cal J}\{f\}$ is the collision integral that accounts for binary collisions (stated in linearized form below). The distribution function $f(t, {\bf r}, {\bf p})$ solving Eq.~\eqref{eq:kinetic} splits into a local equilibrium function $f_0({\bf r}, {\bf p}) = 1/(e^{\beta (\varepsilon - {\bf p} \cdot {\bf V} - \mu)}+1)$ attained by collisions, which moves with center-of-mass velocity ${\bf V}({\bf r})$ ($\beta$~is the inverse temperature and $\mu$ is the chemical potential). For a nonzero velocity, there is a deviation from local equilibrium $\delta f(t, {\bf r}, {\bf p})$ that is induced by the phase-space evolution of single particles. To linear response in the velocity, we expand the local equilibrium function to leading order in the gradient of~${\bf V}({\bf r})$ and determine $\delta f$ from the linearized kinetic equation~\eqref{eq:kinetic},
\begin{equation}
(\partial_x V_y) \, m^* X \biggl(- \frac{\partial f_0}{\partial \varepsilon}\biggr) = {\cal J}\{\delta {f}\} .  \label{eq:deviation}
\end{equation}
The left-hand side originates from the streaming term, which is not annihilated by the local equilibrium distribution, and we abbreviate \mbox{$X = v_x v_y$}. The right-hand side of Eq.~\eqref{eq:deviation} is the linearized collision integral. Defining
\begin{equation}
\delta f(t, {\bf p}) = - (\partial_x V_y) \, m^* \,\Psi(t, {\bf r}, {\bf p}) \biggl(- \frac{\partial f_0}{\partial \varepsilon}\biggr) ,
\label{eq:scalardeformation}
\end{equation}
it is written as
\begin{widetext}
\begin{equation}
\begin{split}
{\cal J}\{\Psi(t, {\bf p})\} &= \frac{2 \pi m^*}{\hbar T} (\partial_x V_y) \int \frac{d{\bf p}_2d{\bf p}_1'd{\bf p}_2'}{(2\pi)^6} \, W({\bf p}, {\bf p}_2; {\bf p}_1', {\bf p}_2') (2\pi)^2\delta({\bf p} + {\bf p}_2 - {\bf p}_{1}' - {\bf p}_{2}')  \, \delta(\varepsilon_{{\bf p}} + \varepsilon_{{\bf p}_2} - \varepsilon_{{\bf p}_{1}'} - \varepsilon_{{\bf p}_{2}'}) \\[1ex]
&\times f_0({\bf p}) f_0({\bf p}_2) \bigl(1-f_0({\bf p}_{1}')\bigr) \bigl(1-f_0({\bf p}_{2}')\bigr)  \bigl[\Psi(t, {\bf p}) + \Psi(t, {\bf p}_2) - \Psi(t, {\bf p}_{1}') - \Psi(t, {\bf p}_{2}')\bigr] .
\end{split} \label{eq:collisionintegral}
\end{equation}
\end{widetext}
This expression describes the change in the quasiparticle distribution due to binary scattering ${\bf p} + {\bf p}_1 \to {\bf p}_1' + {\bf p}_2'$ (and its reverse gain process), which conserves both energy and momentum. Here, for a  spin-independent interaction potential $V({\bf q})$, the scattering matrix element is~\cite{smith89}
\begin{equation}
\begin{split}
W({\bf p}, {\bf p}_2; {\bf p}_1', {\bf p}_2') &= V^2({\bf p}_{1'} - {\bf p}_1) + V^2({\bf p}_{1'} - {\bf p}_2)  \\
&\quad - V({\bf p}_{1'} - {\bf p}_1) V({\bf p}_{1'} - {\bf p}_2) , \label{eq:directexchange}
\end{split}
\end{equation}
which includes both direct (first two terms) and exchange contributions (last term). As mentioned,  the main challenge in obtaining an exact solution of the kinetic Fermi liquid equations is to invert the collision integral~\eqref{eq:collisionintegral} to obtain the deformation function $\Psi$ in response to a shear. 

In terms of the deformation function $\Psi$, the viscosity is compactly expressed as
\begin{equation}
    \eta = \frac{2 m^{*2}}{T \lambda_T^2} \, \langle X | \Psi \rangle ,
\end{equation}
where we introduce the scalar product~\cite{mclennan89,hofmann22b}
\begin{equation}
\langle f | g \rangle = \lambda_T^2 \int \frac{d{\bf p}}{(2\pi)^2} \bigl(- T \frac{\partial f_0}{\partial \varepsilon}\bigr) f^*({\bf p}) g({\bf p}) ,
\label{eq:innerproduct}
\end{equation}
(\mbox{$\lambda_T = \sqrt{2\pi\hbar^2/m^*T}$} is the thermal wave length). The deviation function $\Psi$ that solves the linearized kinetic equation~\eqref{eq:deviation} is compactly expressed as
\begin{equation}
X={\cal L} \Psi \label{eq:deformation}
\end{equation}
if we separate
\begin{equation}
{\cal J}\{\Psi\} = - \bigl(- T \frac{\partial f_0}{\partial \varepsilon}\bigr) {\cal L}\{\Psi\} .
\label{eq:collisionlinear}
\end{equation}
The eigenvalues of the collision integral are the decay rates of different Fermi surface deformations, i.e., they are parameterized by an eigenfunction $\psi_{\alpha}$ that solves \mbox{${\cal L} \psi_{\alpha} = \gamma_{\alpha} \psi_{\alpha}$}~\cite{mclennan89}. To invert the expression~\eqref{eq:deformation} for $\Psi$ and to solve for the viscosity, we expand $\Psi$ in terms of eigenfunctions of the linearized collision integral, \mbox{$\Psi = \sum_{\alpha} c_{\alpha} \psi_{\alpha}$}, such that \mbox{$\langle X | \Psi \rangle = \sum_{\alpha} \gamma_{\alpha} c_{\alpha}^2$} with \mbox{$c_{\alpha} = \langle \Psi | \psi_{\alpha} \rangle = \langle X | \psi_{\alpha} \rangle/ \gamma_{\alpha}$}. It is then straightforward to---at least formally---obtain the viscosity
\begin{equation}
\eta = \frac{2 m^{*2}}{T \lambda_T^2} \, \sum_{\alpha} \frac{|\langle X | \psi_{\alpha} \rangle|^2}{\gamma_{\alpha}} . \label{eq:viscosity}
\end{equation}
This expression is only valid if no zero modes of the collision integral (i.e., with \mbox{$\gamma_{\alpha}=0$}) contribute to the viscosity. Such zero modes are associated with particle and energy conservation [both of which involve a rotationally invariant change of the distribution function, i.e., \mbox{$\psi_\alpha({\bf p}) = \psi_\alpha(|{\bf p}|)$}] as well as current conservation [which involves a rigid shift of the distribution, i.e., \mbox{$\psi_\alpha({\bf p}) = e^{\pm i \theta} \psi_\alpha(|{\bf p}|)$} with $\theta$ the angle of the momentum vector]~\cite{hofmann22b}. From the form of the source \mbox{$X = v_x v_y \sim \sin 2\theta$} and the definition of the inner product (which involves an average over the angle $\theta$), it is clear that only quadrupole perturbations of the form \mbox{$\psi_\alpha \sim e^{\pm2 i\theta}$} contribute to \mbox{$\langle X | \psi_\alpha \rangle$} and hence the viscosity, such that the expression~\eqref{eq:viscosity} is well defined.

The challenge is of course to determine the eigenfunctions $\psi_{\alpha}$. 
We separate the angle dependence of the momentum and write 
\mbox{$\psi_{\alpha\pm} = e^{\pm 2 i \theta} \psi_\alpha^{\pm}(|{\bf p}|)$}, and expand $\psi_\alpha^{\pm}$ in a basis of polynomials $u_\alpha(w)$ in the dimensionless energy variable \mbox{$w = \beta (\varepsilon({\bf p}) - \mu)$}. Starting with a constant function $u_0(w)$, which by Eq.~\eqref{eq:scalardeformation} corresponds to a rigid deformation of the Fermi surface (this is seen by expanding the Fermi-Dirac distribution~\cite{hofmann22}), higher-order polynomials are then generated such that they are orthonormal with respect to the inner product~\eqref{eq:innerproduct}. Note that since the inner product depends on the dimensionless chemical potential $\beta\mu$, the expansion polynomials  are temperature dependent. Once the basis polynomial are determined, it is a numerical task to compute the real-valued matrix elements of the collision integral 
\begin{equation}\label{eq:matrixelement}
\langle u_\alpha(w) | {\cal L} u_{\alpha'}(w) \rangle = -\lambda_T^2 \int \frac{d{\bf p}}{(2\pi)^2} u_\alpha(w) {\cal J}\{u_{\alpha'}(w)\} .
\end{equation}
After solving for energy and momentum conservation, this matrix element involves a higher-dimensional integral with additional kinematic constraints introduced by the Fermi distributions in Eq.~\eqref{eq:collisionintegral}. We evaluate the integrals numerically using the Divonne algorithm for adaptive multidimensional integration contained in the Cuba library~\cite{hahn05}. Including only basis polynomials up to a fixed order ${\cal O}(w^N)$, the computation of the eigenvalues and eigenvectors of ${\cal L}$ then translates to a finite-dimensional matrix eigensystem calculation, where of course it needs to be checked that the result for the shear viscosity converges with increasing basis dimension $N$. Once the eigenvectors are known for a given basis truncation, the matrix elements $\langle X | \psi_\alpha^{(N)}\rangle$, and hence the shear viscosity, follow from an evaluation of the scalar product, which is a simple one-dimensional integral. We note that the formalism outlined and applied in this work is very general and applicable to any linear response function, not just the shear viscosity. 

\begin{figure}[t]
\scalebox{1}{\includegraphics{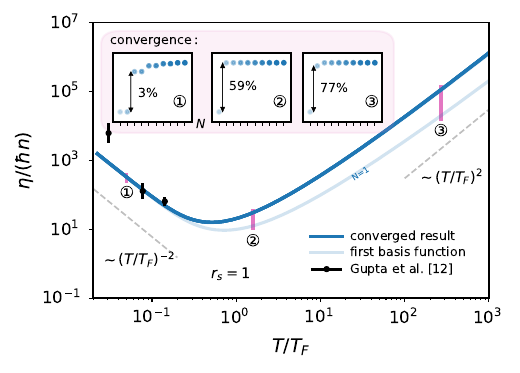}}
\caption{Dimensionless shear viscosity as a function of temperature for a Coulomb interaction strength \mbox{$r_s = 1$}. The continuous blue line shows the converged result extrapolated to infinite basis dimension $N$. The transparent blue line shows the result including only the first basis function (\mbox{$N=1$}), which provides a variational lower bound.
To illustrate convergence, the insets show the viscosity on a linear scale as a function of basis dimension $N$ at selected temperatures indicated by pink markers and numbers in the main plot. Dashed gray lines show the power-law scaling at low and high temperature. Experimental data points are from~\textcite{gupta21}.}
\label{fig:1}
\end{figure}

Figure~\ref{fig:1} shows the result of this calculation for the shear viscosity $\eta$ for an interaction strength \mbox{$r_s=1$} as a function of temperature $T$ in units of the Fermi temperature $T_F = \pi \hbar^2 n / m^* k_B$. The transparent continuous blue line indicates the result using only the first basis function ($N=1$), and the dark blue line shows the converged result as \mbox{$N\to \infty$}. We find that convergence is very rapid at all temperatures, and the quantitative form of the shear viscosity is very similar for all but the very lowest basis dimensions. This is illustrated in the insets, which shows the viscosity as a function of the maximum basis dimension~$N$ on a linear scale for three selected temperatures, the location of which is indicated by pink markers and numbers in the main plot. The insets also include the largest relative jump in precision. These results suggest that already the first two basis functions give an excellent approximation to the full result. Since a fixed basis dimension can be seen as a variational approximation to the shear viscosity, the viscosity in Fig.~\ref{fig:1} strictly increases when enlarging the basis. 

Having established rapid convergence of our solution, we proceed to discuss the parameter dependence of the viscosity. The shear viscosity in Fig.~\ref{fig:1} takes a characteristic form for Fermi systems that changes over many order in magnitude with temperature, with a power-law divergence at both low and high temperatures and a minimum that is attained around $T=0.5 T_F$. At low temperatures, the viscosity diverges as \mbox{$\sim (T_F/T)^2$}, indicative of a characteristic Fermi-liquid time scale $\tau_\eta$. This points to the suppression of interactions by Pauli blocking, with the noninteracting zero temperature gas having infinite viscosity as for a free thermal gas. Right in the hydrodynamic regime where there is a significant broadening of the Fermi surface, the shear viscosity assumes a minimum. Here, our results agree well with the shear viscosity extracted from nonlocal resistance measurements by~\textcite{gupta21} (black markers), which were carried out at densities \mbox{$n\approx 3.4 \cdot 10^{11} {\rm cm}^{-2}$}, corresponding to \mbox{$r_s\approx 1$} (note that at low temperatures, impurity scattering dominates and it is harder to extract the shear viscosity response from the experiment~\cite{gupta21,keser21}). Finally, as the temperature is increased further into the nondegenerate regime \mbox{$n\lambda_T^2 \ll 1$}~\cite{schweng91,hofmann13}, the shear viscosity diverges as \mbox{$\sim (T/T_F)^2$}, which is consistent with the result for the shear viscosity of a classical gas \mbox{$\eta \sim n T \tau_\eta$} with a viscous time scale that is inversely proportional to the scattering cross section, \mbox{$\sigma \sim |V(\lambda_T^{-1})|^2 \sim 1/T$} for a screened Coulomb interaction. Note that for multi-band systems~\cite{principi16} or systems with finite bandwidth~\cite{pakhira15}, the high-temperature region does not have to reach a nondegenerate regime, and the viscosity may saturate with a less pronounced minimum.

\begin{figure}[t]
\scalebox{1}{\includegraphics{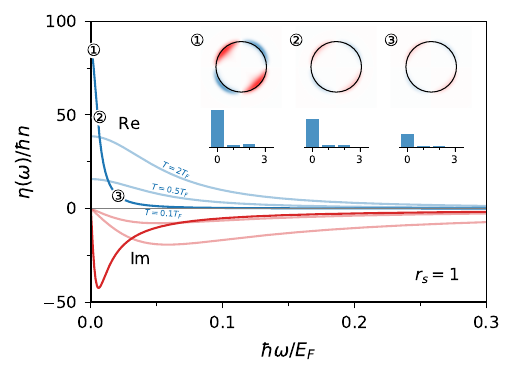}}
\caption{Shear viscosity spectral function $\eta(\omega)$ as a function of frequency for \mbox{$r_s=1$} and three temperatures $T=0.1,0.5$, and $2T_F$. Blue lines indicate the real part and red lines the imaginary part. Top inset panels: Density plot of the quasiparticle deviation $\delta f({\bf p})$ at $T=0.1T_F$ and three frequencies $\hbar\omega/E_F=0,0.025$, and $0.25$, showing excess (blue) and depletion (red) of quasiparticles. Lower inset histograms: Contribution of the basis functions~$u_\alpha$.}
\label{fig:3}
\end{figure}

Results for the viscosity presented so far describe the hydrodynamic response, which relaxes with a characteristic time-scale $\tau_\eta$. A crossover to a collisionless regime with $\omega \tau_\eta \gg 1$ is tuned by the frequency~$\omega$ of the perturbation and described by the complex shear viscosity spectral function $\eta(\omega)$. Such a finite frequency in the kinetic formalism affects the streaming term and changes the correlator~\eqref{eq:viscosity} by \mbox{$\gamma_\alpha \to  \gamma_\alpha - i \omega$} with the same basis functions and matrix elements as for the static viscosity. Figure~\ref{fig:3} shows the resulting spectral function $\eta(\omega)$ at \mbox{$r_s=1$} for three different temperatures $T/T_F=0.1,0.5,$ and $2$, where blue lines indicates the real part and red lines the imaginary part. The spectral function is very narrowly peaked at small and large temperatures where interactions are small and the shear viscosity diverges, but there is significant spectral broadening at intermediate temperatures. The insets of the figure illustrate the Fermi surface deformation~\eqref{eq:scalardeformation} with the solution $\Psi$ at $T=0.5T_F$ for three different frequencies along the hydrodynamic-to-collisionless crossover $\hbar\omega/E_F = 0,0.5$, and $1$, where the upper panels show a density plot of the deformation in momentum space, and the lower panel is a histogram that indicates the strength of the basis functions $\psi_{\alpha}$ for different $\alpha$. The lowest basis function captures the majority of the response of the Fermi gas, consistent with the rapid convergence of the static shear viscosity (cf.~Fig.~\ref{fig:1}).

\begin{figure}[t]
\scalebox{1}{\includegraphics{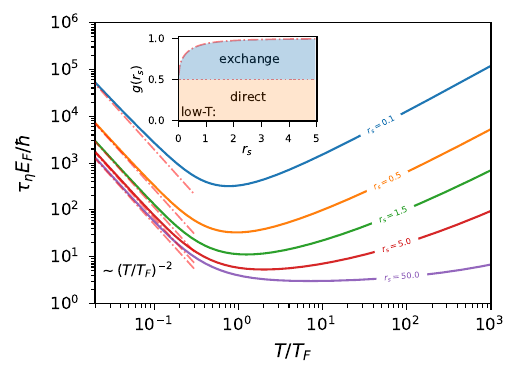}}
\caption{Viscous scattering time $\tau_\eta$ as a function of temperature for \mbox{$r_s=0.1,0.5,1.5,5,$} and $50$. The inset shows the relative direct and exchange contribution at low temperatures as a function $g(r_s)$.}
\label{fig:4}
\end{figure}

At all frequencies, the shear viscosity spectral function is described by the hydrodynamic form~\cite{forster75}
\begin{equation}
\eta_{\rm hydro}(\omega) = \frac{P \tau_\eta}{1+(\omega \tau_\eta)^2} , 
\end{equation}
where the hydrodynamic shear viscosity is related by \mbox{$\eta = P \tau_\eta$}. Here, $P$ is the Fermi gas pressure, which is constant \mbox{$P=nE_F/2 = nmv_F^2/4$} at low temperatures and changes to the linear ideal gas pressure \mbox{$P=(T/T_F)nE_F$} at high temperatures. The viscous scattering time $\tau_\eta$ sets the relaxation time for Fermi surface deformations in response to a velocity shear and is determined by collisions. Figure~\ref{fig:4} shows this quantity as a function of temperature for five different interaction strengths $r_s=0.1,0.5,1.5,5$, and $50$. It follows Fermi-liquid scaling \mbox{$\tau_\eta \sim (T_F/T)^2$} at low temperatures and crosses over to a linear temperature dependence for a classical gas as discussed above. While the position of the minimum of $\tau_\eta$ increases weakly with the interaction strength (for realistic values of $r_s$), the onset of the crossover from the low-temperature region does not, which indicates that the minimum is linked to the softening of the Fermi-Dirac distribution with increasing temperature that weakens the Pauli blocking. Correspondingly, we find that the position of the minimum of the shear viscosity depends only weakly on  interaction strength. This shear viscosity minimum is seen as a measure of how close the the fluid comes to a ``perfect'' fluid with a conjectured bound (adapted to non-relativistic systems) \mbox{$\eta/n \geq \hbar/(4\pi)$} from high-energy physics~\cite{kovtun05}, which has been checked in experiments on heavy-ion collisions~\cite{heinz13} as well as strongly interacting ultracold quantum gases~\cite{cao11,cao11b}. For the electron gas, we find that the minimum of \mbox{$\eta/\hbar n$} is at least a factor $50$ larger than this bound. Saturating the bound would imply a breakdown of the quasiparticle description~\cite{danielewicz85,schaefer14}, which points to the consistency of our Fermi-liquid description at all temperatures.

Including only the lowest basis function $u_0$ in the matrix element~\eqref{eq:matrixelement}, we obtain an analytical result for the viscous scattering time at low temperatures,
\begin{equation}
\tau_\eta = \frac{3\hbar}{2 \pi T_F} \frac{T_F^2}{T^2} r_s^{-2} \biggl[ \ln \Bigl( 1 + \frac{\sqrt{2}}{r_s}\Bigr) - \frac{\sqrt{2}}{\sqrt{2}+r_s}\biggr]^{-1} g(r_s),
    \label{eq:lowT_even_pred}
\end{equation}
which we show as red dashed-dotted lines in Fig.~\ref{fig:4}. We separate a function $g(r_s)$ (shown in the inset of Fig.~\ref{fig:4}) that describes the relative contribution of the direct and exchange scattering to the shear viscosity. From Eq.~\eqref{eq:directexchange}, the exchange contribution to the scattering rate $\tau_\eta^{-1}$ subtracts at most one half of the direct contribution, doubling the shear viscosity. For most $r_s$, the exchange contribution is indeed maximal and \mbox{$g(r_s)=1$}, and is only suppressed below very small values $r_s\leq 0.1$ with a nonanalytic dependence on $r_s$. For very small $r_s$, we have \mbox{$(T_F/\hbar)\tau_\eta = 3 (T_F/T)^2/(4\pi r_s^2 \ln(\sqrt{2}/r_s e^{1+\pi/4}))$}, and at large~$r_s$, we have a constant \mbox{$(T_F/\hbar)\tau_\eta = 3 (T_F/T)^2/(2\pi)$}. The small-$r_s$ limit agrees with an analytical calculation~\cite{alekseev20} (note that our expression for the amplitude~\eqref{eq:directexchange} differs from~\cite{alekseev20} by a factor of 2), but due to the rapid nonanalytic increase in the exchange contribution its validity is restricted to asymptotically small~$r_s$. This indicates that established methods to analyze the 3D Fermi gas~\cite{abrikosov57, abrikosov59} do not describe the 2D systems beyond the asymptotic weak-interaction regime. Moreover, there is also no correction to Eq.~\eqref{eq:lowT_even_pred} that is logarithmic in temperature, which is the canonical result for the quasiparticle lifetime~\cite{giuliani82}. A simple relaxation time approximation is thus also inaccurate to predict the shear viscosity transport time scale $\tau_\eta$ even at small temperature~\cite{principi16}.
 
In summary, we have established a numerically exact solution of the Fermi liquid equations to compute the shear viscosity in an interaction-dominated electron gas. Such calculations are pertinent given recent experimental progress to perform high-precision tests of Fermi-liquid many-body physics at finite temperatures, and necessary because standard many-body techniques such as the relaxation-time approximation and leading-logarithmic approximations have only limited range of validity even at low temperatures. The formalism outlined here is applicable to the general computation of transport coefficients in Fermi systems and valid in the full crossover from a degenerate to a nondegenerate gas. For the case of the shear viscosity, our framework can serve as a platform to investigate the effect of electron correlations in the interaction matrix element beyond the simple Coulomb interaction used here~\cite{keser21} or additional interaction contributions to the viscosity~\cite{liao20}, and to include the coupling to bosonic modes such as phonons or plasmons~\cite{pongsangangan23,zverevich23}. 

\begin{acknowledgments}
This work is supported by Vetenskapsr\aa det (Grant No. 2020-04239) (JH) and the Chalmers' Area of Advance Nano under its Excellence Ph.D.\@ program (EN).
\end{acknowledgments}

\bibliography{bib_shear}

\end{document}